\documentclass[letter]{jpconf}

\usepackage{graphicx}
\usepackage{epstopdf}

\begin{document}


\title{Cosmic Neutrino Bound on the Dark Matter Annihilation Rate in the Late Universe}

\author{John F. Beacom}

\ead{beacom@mps.ohio-state.edu}

\address{Department of Physics, Department of Astronomy,
and  Center for Cosmology and Astro-Particle Physics,
The Ohio State University, Columbus, Ohio 43210}

\begin{abstract}
How large can the dark matter self-annihilation rate in the late universe be?
This rate depends on $(\rho_{DM}/m_\chi)^2 \langle \sigma_A v \rangle$,
where $\rho_{DM}/m_\chi$ is the number density of dark matter, and the
annihilation cross section is averaged over the velocity distribution.
Since the clustering of dark matter is known, this amounts to asking how 
large the annihilation cross section can be.  Kaplinghat, Knox, and Turner
proposed that a very large annihilation cross section could turn a halo cusp
into a core, improving agreement between simulations and observations;
Hui showed that unitarity prohibits this for large dark matter
masses.  We show that if
the annihilation products are Standard Model particles, even just neutrinos,
the consequent fluxes are ruled out by orders of magnitude, even at small
masses.  Equivalently, to invoke such large annihilation cross sections, one
must now require that essentially no Standard Model particles are produced.

\medskip
(This proceedings article is based on J.~F.~Beacom, N.~F.~Bell and G.~D.~Mack,
``General Upper Bound on the Dark Matter Total Annihilation Cross Section,''
astro-ph/0608090~\cite{fullpaper}.)
\end{abstract}


\section{Dark Matter Disappearance}

The self-annihilation cross section is a fundamental property of dark
matter.  For thermal relics, it sets the dark matter mass
density, $\Omega_{DM} \simeq 0.3$, and in these and
more general non-thermal scenarios, also
the annihilation rate in gravitationally-collapsed dark matter
halos today~\cite{BHS}.
There are two general constraints that bound the rate of dark matter
{\it disappearance}.  (Throughout, we mean the cross section
averaged over the halo velocity distribution, i.e., $\langle \sigma_A
v \rangle$, where $v_{rms} \sim 10^{-3}c$.)  We assume that the dark matter
is its own antiparticle.

The first is the unitarity bound, developed for the early universe
case by Griest and Kamionkowski~\cite{Griest}, and for the
late-universe halo case by Hui~\cite{Hui}.  In the plane of $\langle
\sigma_A v \rangle$ and dark matter mass $m_\chi$, this allows only
the region below a line $\langle \sigma_A v \rangle \sim 1/m_\chi^2$.
The unitarity bound has
some possible exceptions~\cite{Griest, Hui, Kusenko}.
The second is provided by the
model of Kaplinghat, Knox, and Turner (KKT)~\cite{KKT}, in which
significant dark matter annihilation is invoked to resolve a
conflict between predicted (sharp cusps) and observed (flat cores)
halo profiles.  (In contrast to using dark matter elastic self-scattering,
as in Ref.~\cite{Spergel}.)  Since this tension may have been
relaxed~\cite{BHS}, we reinterpret this type of model as an upper
bound, allowing only the region below a line $\langle \sigma_A v
\rangle \sim m_\chi$.  That the KKT model requires $\langle \sigma_A v
\rangle$ values $> 10^7$ times larger than the natural scale for
a thermal relic highlights the weakness of the unitarity bound in the
interesting GeV range.  However, there have been no other strong and
general bounds to improve upon these.   (Following Refs.~\cite{Hui, KKT},
we assume that the dark matter is somehow not a thermal relic, and
hence ignore early-universe constraints, which would prohibit such large
cross sections.)

While these bound the {\it disappearance} rate of dark matter, they
say nothing about the {\it appearance} rate of annihilation products,
instead assuming that they can be made undetectable.  To
evade astrophysical limits, the branching ratios to specific final
states can be adjusted in model-dependent ways.  However, a
model-independent fact is that the branching ratios for all final
states must sum to 100\%.  A reasonable assumption
is that these final states are Standard Model (SM) particles, as it is
assumed the dark matter is the lightest stable particle in the
Beyond-SM sector; we generalize below.  


\section{Neutrino Appearance}

We assume that annihilation proceeds to SM particles, and
express the cross section in terms of branching ratios
to ``visible'' and ``invisible'' final states, such as gamma rays and
neutrinos, respectively.
If the branching ratio to a specific final state were known, then a
bound on that appearance rate would yield a bound
on the total cross section, inversely proportional to this branching
ratio.  However, the branching ratios are model-dependent, and
any specific one can be made very small, making that bound on
$\langle \sigma_A v \rangle$ very weak, e.g., KKT require a stringent
$Br(\gamma) < 10^{-10}$ to avoid exceeding
the cosmic diffuse gamma-ray background~\cite{KKT}.

KKT~\cite{KKT} and Hui~\cite{Hui} assume invisible final states,
but do not specify them.  Among
SM final states, it is clear that all but neutrinos
will produce many more gamma rays than $Br(\gamma) \sim 10^{-10}$.
Quarks and gluons hadronize, unavoidably producing pions, where
$\pi^0 \rightarrow \gamma \gamma$; the decays of weak bosons and tau
leptons also produce $\pi^0$.
It has been shown recently that the stable final state $e^+ e^-$ is not
invisible, since it produces gamma rays either through
electromagnetic radiative corrections~\cite{RC} or energy loss
processes~\cite{Eloss}; the final state $\mu^+ \mu^-$ immediately
produces $e^+ e^-$ by its decays.
Thus the only possible ``invisible" SM final states are neutrinos.
Of final states
with only neutrinos, we focus on $\bar{\nu} \nu$.  Similar bounds
could be derived for $\bar{\nu} \bar{\nu} \nu \nu$, but we assume that
these are either suppressed and/or that the Rube Goldberg-ish Feynman
diagrams required would contain charged particles, and hence gamma
rays through radiative corrections.

\begin{figure}
\includegraphics[width=18pc]{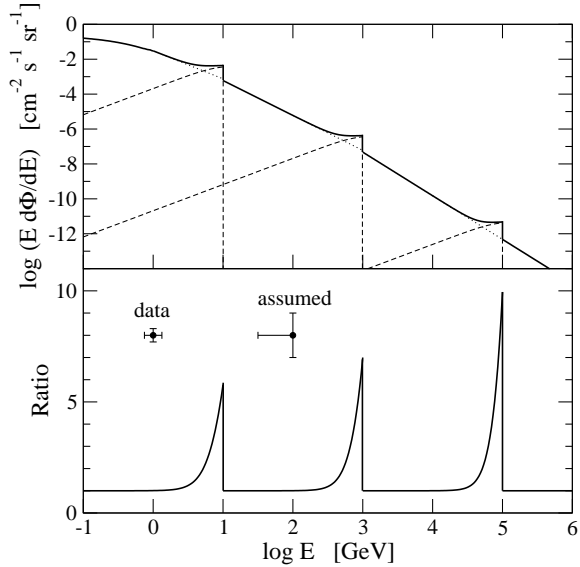}
\hspace{2pc}
\begin{minipage}[b]{18pc}
\caption{\label{fig:spectrumplot}
{\bf Upper:} Diffuse $\bar{\nu} \nu$ annihilation signal for
$m_\chi = 10, 10^3,$ and $10^5$ GeV, added to the atmospheric
background, both as ($\bar{\nu}_\mu + \nu_\mu$) and versus neutrino
energy.  As noted, the signals are most accurate for $E_\nu >
m_\chi/3$.  {\bf Lower:} Ratio of this sum and background.  The
$\langle \sigma_A v \rangle$ values at each example $m_\chi$ are
chosen to be detectable by our conservative criteria; the data and
assumed uncertainty scales are also indicated.
\\
Figure taken from Ref.~\cite{fullpaper}.}
\end{minipage}
\end{figure}

To derive our bound on the {\it total}
annihilation cross section, we assume $Br(\bar{\nu} \nu) = 100\%$.
This is {\it not} an assumption about realistic outcomes, but it is the
right way to derive the {\it most conservative} upper bound,
if only SM final states are possible.
Why is this a bound on the total cross section, and not just on the
partial cross section to neutrinos?  Because if even a small
fraction of the final states were not neutrinos, they would produce
gamma rays, and those flux bounds are so much more stringent
that the assumed cross section would be ruled out.
Therefore, while setting this bound using neutrinos can be too
conservative, it can never overreach.
It is beyond our scope to set a more stringent bound using either
direct gamma rays (probing $\langle \sigma_A v \rangle$
near the natural scale~\cite{BHS}), or those produced via radiative
corrections~\cite{Berezinsky}; this would require specifying the
branching ratios for these final states.


\section{Neutrino Signal and Background}

\begin{figure}
\includegraphics[width=18pc]{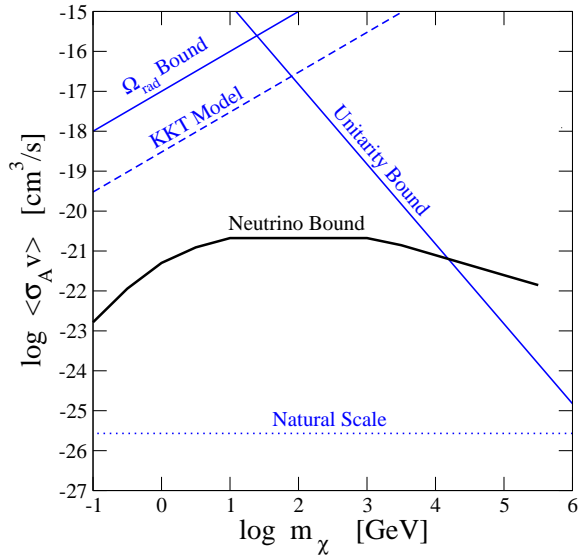}
\hspace{2pc}
\begin{minipage}[b]{18pc}
\caption{\label{fig:sigmavplot}
Upper bounds on the dark matter total annihilation cross section in
galaxy halos as a function of the dark matter mass, calculated as discussed
in the text.
\\
Figure taken from Ref.~\cite{fullpaper}.}
\end{minipage}
\end{figure}

To bound the neutrino appearance rate, we use
the cosmic diffuse neutrino flux from
dark matter annihilations in all halos in the universe as the signal.
Since this is isotropic and time-independent, it is challenging to
detect above the background caused by the atmospheric
neutrino flux.  Our calculations depend on the average value of
density squared in the late (clustered) universe, and for this, we
follow the calculations of Refs.~\cite{UllioPRD, UllioPRL, others}.
The formalism is summarized in Ref.~\cite{fullpaper}, along with
a detailed discussion of the atmospheric neutrino
backgrounds~\cite{SK-GC, Frejus, AMANDA, SK, SK-HE, AtmNu}
and how large of a signal perturbation is allowed.

The predicted neutrino spectra are shown in Fig.~\ref{fig:spectrumplot}.
These results can be simply checked, following some
basic principles.  We plotted our spectra as $E d\Phi/dE$ to make these
points obvious, and to make estimating the integrals over energy a
trivial multiplication by $d\log E$.  First, since the annihilation rate
scales as $n^2 = (\rho_{DM}/m_\chi)^2$, the energy-integrated fluxes should scale
as $1/m_\chi^2$, and they do.  Second, since the redshift history is
independent of dark matter mass, the spectral shapes for different
masses should be the same ($E d\Phi/dE$ at each energy is proportional to
the number of annihilations at the corresponding redshift), up to the
$1/m_\chi^2$ normalization above, and they are.  Third, the fluxes should be
dominated by annihilation at low redshift, and they are.  Fourth, one
can estimate the energy-integrated fluxes by multiplying the
annihilation rate density by several Gyr and $c/4\pi$, and this comes out right.


\section{Conclusions}

We have shown that the dark matter total annihilation cross section
in the late universe,
i.e., the dark matter {\it disappearance} rate, can be directly and
generally bounded by
the least detectable SM states, i.e., the neutrino {\it appearance} rate.
This can be simply and robustly constrained by comparing the diffuse
signal from all dark matter halos to the terrestrial
atmospheric neutrino background.  Our final bound on $\langle \sigma_A v
\rangle$ is shown in Fig.~\ref{fig:sigmavplot}.
Over a very large range in $m_\chi$, 
it is much stronger than the
unitarity bound of Hui~\cite{Hui}.  It strongly rules out the proposal of
Kaplinghat, Knox, and Turner~\cite{KKT} to modify dark matter halos by
annihilation.  The only exception unique to our bound is if one
postulates truly invisible non-SM final states, such as sterile neutrinos.
However, the annihilations would have to proceed to {\it only} those states.
For a cross section
above our bound, its ratio to our bound yields a conservative
constraint on the branching ratio to SM final states that one would
have to require to invoke that cross section.

Annihilation flattens halo cusps to a core of density
$\rho_A \sim m_\chi/(\langle \sigma_A v \rangle H_0^{-1})$~\cite{KKT}.
Our bound implies that for all $m_\chi > 0.1$ GeV, this density
is $\rho_A > 5 \times 10^3 {\rm\ GeV/cm}^3$, which only occurs
at radii $< 1 $ pc in the Milky Way for an NFW profile,
and perhaps not at all for less steep profiles.  While modeling is needed
to confirm this, we expect that dark matter annihilation
cannot have any macroscopic effect on galactic halos, and that it is
unlikely that such a strong statement could be made any other way
(for example, an argument that we made based on the radiation
density~\cite{Zentner} is too weak).
Detailed analyses by the Super-Kamiokande and AMANDA Collaborations
should be able to improve our bound by a
factor 10--100 over the whole mass range.  Halo substructure or
mini-spikes around intermediate-mass black holes could increase the
signal by orders of magnitude~\cite{others, MoreSignal}.  The
sensitivity could thus become close to the natural scale for thermal
relics, making it a new tool for testing even standard scenarios.


\ack
This proceedings article is based on collaborative work~\cite{fullpaper}
with Nicole Bell and Greg Mack, whom I gratefully acknowledge.
We are grateful to Gianfranco Bertone for very helpful discussions
and collaboration on an early stage of this project.  We thank
Shin'ichiro Ando, Shantanu Desai, Michael Kachelriess, Manoj
Kaplinghat, Eiichiro Komatsu, Bob Scherrer, and Hasan Y\"uksel
for helpful discussions.
JFB was supported by NSF CAREER Grant No. PHY-0547102,
NFB by a Sherman Fairchild fellowship at Caltech,
and GDM by DOE Grant No. DE-FG02-91ER40690;
we also thank CCAPP and OSU for support.



\end{document}